\documentclass[letter]{ptptex}
\usepackage[dvips]{graphicx,color}
\usepackage{multirow}
\usepackage{bm}
\usepackage{wrapft}

\def\dfrac#1#2{{\displaystyle\frac{#1}{#2}}}

\def\beq{\begin{equation}}
\def\eeq{\end{equation}}
\def\bea{\begin{eqnarray}}
\def\eea{\end{eqnarray}}

\def\la{\mathrel{\mathpalette\fun <}}
\def\ga{\mathrel{\mathpalette\fun >}}
\def\fun#1#2{\lower3.6pt\vbox{\baselineskip0pt\lineskip.9pt
\ialign{$\mathsurround=0pt#1\hfil##\hfil$\crcr#2\crcr\sim\crcr}}}
\title{
Quantum three-body calculation of
the nonresonant triple-$\alpha$ reaction rate
at low temperatures
}

\author{
Kazuyuki \textsc{Ogata}$^{1}$\thanks{E-mail: ogata@phys.kyushu-u.ac.jp},
Masataka \textsc{Kan}$^{1}$\thanks{Present address: Advanced Information
Systems R\&D Department, Hitachi Ltd., Kawasaki 212-8567, Japan},
and
Masayasu \textsc{Kamimura}$^{1,2}$
}

\inst{
$^{1}$Department of Physics, Kyushu University,
Fukuoka 812-8581, Japan \\
$^{2}$RIKEN Nishina Center, 351-0198 Wako, Japan
}

\abst{
The triple-$\alpha$ reaction rate is re-evaluated
by directly solving the three-body Schr\"{o}dinger equation.
The resonant and nonresonant processes are treated on the same footing
using the continuum-discretized coupled-channels method for
three-body scattering.
Accurate description of the $\alpha$-$\alpha$ nonresonant states
significantly quenches the Coulomb barrier between the two-$\alpha$'s
and the third $\alpha$ particle.
Consequently, the $\alpha$-$\alpha$ nonresonant continuum states below the
resonance at 92.04~keV, i.e., the ground state of $^8$Be,
give markedly larger contribution
at low temperatures
than in foregoing studies.
We show that Nomoto's method for three-body
nonresonant capture processes,
which is adopted in the NACRE compilation and many other studies,
is a crude approximation of the accurate quantum three-body model calculation.
We find about 20 orders-of-magnitude
enhancement of the triple-$\alpha$ reaction rate around $10^7$~K
compared to the rate of NACRE.
}

\begin{document}
\maketitle

After the Big Bang, all elements surrounding us have been
created in the cosmos. Among them, $^{12}$C is one of the most
important nuclei because it is an essential element of life.
In this sense, to understand the origin of $^{12}$C would be
equivalent to know {\it {\lq\lq}where we come from.''}
In view of nuclear physics, it is well known that
the second $0^+$ state of $^{12}$C at 7.65~MeV above its
ground state was predicted by Hoyle in 1953
to explain the abundance of $^{12}$C, or, existing
of human beings; this challenging prediction
was soon confirmed experimentally~\cite{Dunbar}, and the
$0^+_2$ state newly discovered is called the Hoyle resonance.
Since this historic discovery of the Hoyle resonance,
the $\alpha(\alpha\alpha,\gamma)$$^{12}$C reaction, i.e., the
so-called triple-$\alpha$
reaction, has been described as a series of the following
two reactions~\cite{Rolfs}:
\begin{equation}
\alpha+\alpha \rightarrow {\rm ^{8}Be},
\quad
\alpha+{\rm ^{8}Be} \rightarrow {\rm ^{12}C}(2^+_1)+\gamma.
\label{reac}
\end{equation}
In the latter, $^{12}$C in the $2^+_1$ state is formed by a $\gamma$-decay
from the Hoyle resonance, and then decays into the ground state
of $^{12}$C, with emitting $\gamma$ again.
This picture of the triple-$\alpha$
reaction, which is realized at temperatures higher than
a few $10^8$~K in helium burning stars,
successfully describes the present abundance of $^{12}$C.

At low temperatures, however,
the energy of the three $\alpha$ system cannot reach the resonance
energy of 379.5~keV above the three-$\alpha$ threshold.
A formation process through nonresonant three-$\alpha$
continuum states, i.e., the nonresonant triple-$\alpha$ process,
then becomes dominant.
The formation rate of $^{12}$C at low temperatures
is of crucial importance for
studies on helium burning in accreting white
dwarfs and neutron stars~\cite{Nomoto1,Nomoto2},
and is believed to strongly affect the evolution of primordial
stars~\cite{Nature}. \
In Ref.~\citen{Nomoto1} Nomoto proposed a method for evaluating
contribution of the nonresonant triple-$\alpha$ process
by using resonance formulae, with an {\it energy shift}
for the Hoyle resonance, to describe the two reactions of Eq.~(\ref{reac}).
This method, which we call Nomoto's method in this Letter,
was found to give a significantly larger triple-$\alpha$ reaction rate
at low temperatures~\cite{Nomoto1,Nomoto3}
than that obtained by a naive resonance formula without
the energy shift.
Nomoto's method has been a standard method for describing the nonresonant
triple-$\alpha$ reaction. It should be noted, however, that
Nomoto's method does not explicitly
describe the role of the nonresonant continuum states in
the triple-$\alpha$ process.

In the present Letter, we evaluate the nonresonant
triple-$\alpha$ reaction rate
by directly solving the Schr\"{o}dinger equation of
the three-$\alpha$ system.
The three-$\alpha$ scattering wave function is
obtained by the continuum-discretized coupled-channels method
(CDCC)~\cite{CDCC1}, which was proposed and developed by
the Kyushu group more than 20 years ago, and
has been successfully applied to studies of various
three-body reaction processes; see, e.g., Refs.~\citen{CDCC1,Ogata,Surrey}.
In CDCC, resonant and nonresonant
states of the $\alpha$-$\alpha$ system,
as well as those of the three-$\alpha$ system,
are treated on the same footing.
This is one of the most important advantages of the present calculation.

As for previous three-body calculations of the
triple-$\alpha$ reaction, Kamimura and Fukushima~\cite{Kamimura1}
firstly showed, using the microscopic three-$\alpha$ resonating group
method, the importance of the couplings between the $\alpha$-$\alpha$
resonant channel and the $\alpha$-$\alpha$
nonresonant channels for reproducing
the properties of the Hoyle resonance and understanding the
processes of Eq.~(\ref{reac}). This finding was confirmed by
Descouvemont and Baye~\cite{DB} using the microscopic
three-$\alpha$ generator coordinate method; they derived
the triple-$\alpha$ reaction rate down to $10^7$~K.
In these studies,
however, treatment of the $\alpha$-$\alpha$ nonresonant continuum
was very primitive; only a few discretized nonresonant states
were included.
More seriously, $\alpha$-$\alpha$ nonresonant continuum states below the
resonance at 92.04~keV, which play essential roles in
the nonresonant triple-$\alpha$ process as shown below,
were completely missed. Thus, it is obvious that these models cannot
accurately describe
the nonresonant triple-$\alpha$ process at low temperatures,
say, $T \la 10^8$~K.

CDCC is well known as one of the most accurate
reaction models to describe
three-body reactions. It works even cases where
Coulomb interactions play dominant roles
and careful description of low-energy continuum states
is required.
In the present study, we use the three-$\alpha$ scattering
wave function obtained by CDCC, and
evaluate the triple-$\alpha$ reaction rate
at temperatures of $10^7$~K $\le T \le 10^9$~K.
Emphasis is on the reaction rate at low temperatures for
$T \le 10^8$~K, where contribution of the nonresonant triple-$\alpha$
process is dominant.
We show that Nomoto's method~\cite{Nomoto1,Nomoto3}
used in the NACRE compilation~\cite{NACRE}
and many other studies is a crude approximation of the accurate
quantum three-body model calculation.
At $T=10^7$~K, typically, our new result is more than
20 orders-of-magnitude as large as that of NACRE~\cite{NACRE}.

%
\begin{figure}[htbp]
\begin{center}
\includegraphics[width=0.45\textwidth,clip]{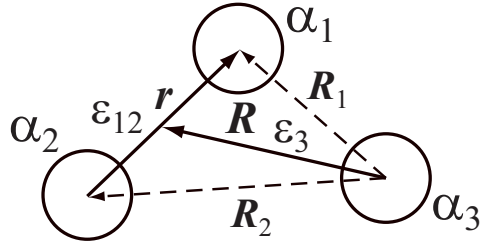}
\caption{Illustration of the three-$\alpha$ system.}
\label{fig1}
\end{center}
\end{figure}
In the present calculation, we work with the three-$\alpha$ system
shown in Fig.~\ref{fig1}
based on the Jacobi coordinates (${\bm r}$, ${\bm R}$).
The relative energy
of $\alpha_1$ and $\alpha_2$ is denoted by $\epsilon_{12}$
and the relative energy between
$\alpha_3$ and the center-of-masses (c.m.) of $\alpha_1$ and $\alpha_2$
is denoted by $\epsilon_3$; the total energy $E$
in the three-$\alpha$ c.m. frame is $\epsilon_{12}+\epsilon_3$.
We obtain the triple-$\alpha$ reaction rate by directly solving the following
three-body Schr\"{o}dinger equation:
\begin{equation}
\left[
T_{\bm r}+T_{\bm R}
+v(r)
+v(R_1)
+v(R_2)
-E
\right]
\Psi ({\bm r},{\bm R})=0,
\label{SchEq}
\end{equation}
where $T_{\bm r}$ and $T_{\bm R}$ are the kinetic energy operators
associated with ${\bm r}$ and ${\bm R}$, respectively, and
$v$ is the interaction between two-$\alpha$'s
consisting of the nuclear and Coulomb parts.

Following the standard continuum-discretization procedure called
{\it the average method} \cite{CDCC1},
we first discretize the continuum states of the
$\alpha_1$-$\alpha_2$ subsystem into momentum bins.
We prepare $L^2$-integrable
$\alpha_1$-$\alpha_2$ wave functions $\hat{u}_{i}(r)$
($i=1$--$i_{\rm max}$) by
\begin{equation}
\hat{u}_{i}(r)=
\int_{k_{i}}^{k_{i+1}} f_i(k) u(k,r) dk
\bigg/
\left[
\int_{k_{i}}^{k_{i+1}} |f_i(k)|^2 dk
\right]^{1/2},
\label{uhat}
\end{equation}
where $u(k,r)$ is the $\alpha_1$-$\alpha_2$ scattering wave function
with the relative momentum $k$, and $f_i(k)$ is a weight function
set to be a constant and the Breit-Wigner function
for nonresonant and resonant bin states, respectively~\cite{CDCC1}. \
Normalization of $u(k,r)$ is defined by
$\int u^*(k',r) u(k,r) dr = \delta (k'-k)$, which makes
$\hat{u}_{i}(r)$ satisfy
$\int \hat{u}^*_{i}(r) \hat{u}_{j}(r) dr =\delta_{ij}$.
The average momentum (energy)
of the $i$th {\it bin} state of the $\alpha_1$-$\alpha_2$ system
is denoted by $\hat{k}_{i}$ ($\hat{\epsilon}_{12,i}$) in the following.
The total wave function of the three-$\alpha$ system with CDCC
is given by
\[
\Psi_{\hat{k}_{i_0},E}^{0^+}\left(  r,R\right)
=
\sqrt{\frac{2}{\pi}}\frac{1}{32\pi^{2}}
\frac{1}{\hat{k}_{i_0}\hat{K}_{i_0}}
\sum_{i=1}^{i_{\rm max}}
\frac{\hat{u}_{i}\left(  r\right)  }{r}
\frac{\hat{\chi}_{i}^{(i_0)}(R)  }{R},
\]
where $\hat{\chi}_{i}^{(i_0)}(R)$ describes the
relative motion between ($\alpha_1$-$\alpha_2$) in
$\hat{u}_{i}$ and $\alpha_3$
with the relative momentum $\hat{K}_{i}$ that
is obtained by energy conservation of the three-$\alpha$ system.
The index $i_0$ represents the incident channel.
Note that we consider only the s-waves of $\hat{u}_{i}$ and
$\hat{\chi}_{i}^{(i_0)}(R)$, since we are interested in a reaction
at very low energies.
Furthermore, we have dropped
all phase factors in $\Psi_{\hat{k}_{i_0},E}^{0^+}$
giving no contribution to the reaction probability shown below.
The coupled-channel (CC) equations for $\hat{\chi}_{i}^{(i_0)}(R)$
($i=1$--$i_{\rm max}$) are given by
\begin{equation}
\left[
T_{R}
+ V_{ii} \left(  R\right)
- \left(  E - \hat{\epsilon}_{12,i}\right)
\right]  \hat{\chi}_{i}^{(i_0)}(R)
= -  \sum_{i'\neq i}
V_{ii'} \left(  R\right)
\hat{\chi}_{i'}^{(i_0)}(R),
\label{CDCCeq}
\end{equation}
which are solved under a usual boundary condition for
$\hat{\chi}_{i}^{(i_0)}(R)$~\cite{CDCC1}. \
The coupling potential $V_{ii'}(R)$ is defined by
\begin{equation}
V_{ii'}\left(  R\right)  =\left\langle
\frac{\hat{u}_{i}\left(  r\right)  }{r}
\bigg\vert
v\left(R_{1}\right)
+v\left(  R_{2}\right)
\bigg\vert
\frac{\hat{u}_{i'}\left(  r\right)  }{r}
\right\rangle_{\bm r}.
\label{FF}
\end{equation}

%
\begin{figure}[b]
\begin{center}
 \includegraphics[width=0.55\textwidth,clip]{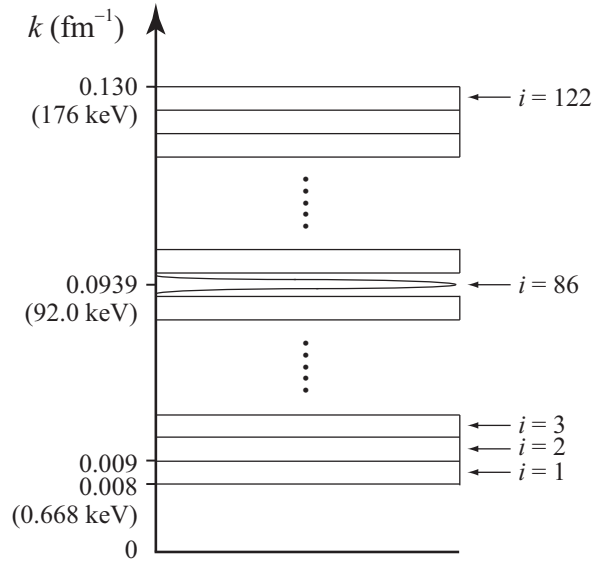}
 \caption{
 $\alpha_1$-$\alpha_2$ momentum bin states included in the present
 CDCC calculation.}
\label{fig2}
\end{center}
\end{figure}
The reaction probability
$\left\langle \sigma v \right\rangle_{\hat{k}_{i_0},E}$
of the triple-$\alpha$ process due to the
electric quadrupole (E2) transition to the $2_1^+$ state
is given by
\begin{equation}
\left\langle \sigma v \right\rangle_{\hat{k}_{i_0},E}
=
\frac{2\left(2\pi\right)^{7}}{75\hbar}\left(  \frac{\hbar\omega}{\hbar
c}\right)  ^{5}
\sum_{M}\left\vert \left\langle
\Psi_{M}^{2^+}\left\vert O^{\rm E2}_{M}  \right\vert
\Psi_{\hat{k}_{i_0},E}^{0^+} \right\rangle \right\vert ^{2},
\nonumber
\end{equation}
where $\Psi_{M}^{2^+}$ is the wave function of the $2_1^+$ state
of $^{12}$C with $M$ the projection of the total spin,
$O^{\rm E2}_{M}$ is the E2 transition operator,
and the photon energy $\hbar \omega$ is given by
$2.8358+E$~MeV.
Note that we use the symbol
$\left\langle \sigma v \right\rangle$
for the reaction probability following NACRE~\cite{NACRE};
it is actually the E2 transition
probability divided by the normalization factor $1/(2\pi)^6$
for the three-$\alpha$ scattering wave function in free space.
The triple-$\alpha$ reaction rate
$\langle \alpha\alpha\alpha\rangle (T)$, as a function of
$T$, is obtained by taking an average of
$\left\langle \sigma v \right\rangle_{\hat{k}_{i_0},E}$
with the Maxwell-Boltzmann distribution for the velocity
of each $\alpha$:
\begin{equation}
\langle \alpha\alpha\alpha\rangle (T)
=
3 N_{\rm A}^{2}
\frac{4}{\pi\left(  k_{\rm B}T\right)  ^{3}}
\int\left\{  \sum_{i_0=1}^{i_{\mathrm{\max}}}w_{i_0}\left\langle
\sigma v \right\rangle_{\hat{k}_{i_0},E} \right\}
\exp\left(  -\frac{E}{k_{\rm B}T}\right)  dE
\label{rate}
\end{equation}
with
\[
w_{i_0}=\dfrac{2\hat{\epsilon}_{12,i_0}}{\hat{k}_{i_0}}
\sqrt{\hat{\epsilon}_{12,i_0}(E-\hat{\epsilon}_{12,i_0})},
\]
where $k_{\rm B}$ is Boltzmann's constant and $N_{\rm A}$
is Avogadro's number.
Factor 3 in Eq.~(\ref{rate}) comes from
the fact that the three-$\alpha$ system is symmetric with respect to
an exchange of each pair of alpha particles~\cite{NACRE}.

In numerical calculation, we discretize
the $k$-continuum of the
$\alpha_1$-$\alpha_2$ system from
0.008 fm$^{-1}$ ($\epsilon_{12}=0.668$~keV)
to 0.130 fm$^{-1}$ ($\epsilon_{12}=176$~keV) with the width of
0.001 fm$^{-1}$, which results in $i_{\rm max}=122$.
Schematic illustration of the $\alpha_1$-$\alpha_2$
bin states taken in the CDCC calculation
is shown in Fig.~\ref{fig2}.
This discretization is sufficiently precise for the present purpose.
The 86th bin corresponds to the $\alpha_1$-$\alpha_2$ resonance.
The maximum value of $r$ is 5,000 fm with the increment $\Delta r$ of 0.1 fm.

As for the nuclear potential between two $\alpha$'s,
we use the following two-range Gaussian form (with depth in unit of MeV):
\begin{equation}
v^{\rm nucl}(x)
= 100.0 e^{-(x/1.00)^2} - 30.35 e^{-(x/2.13)^2},
\label{Vaa}
\end{equation}
where $x$ is the displacement of the two $\alpha$'s in fm. The first repulsive
part simulates the Pauli exclusion principle that nucleons in an $\alpha$
cannot occupy the nucleon s-orbit in the other.
This potential gives $\alpha$-$\alpha$ resonance at
92.0~keV with the width of 4.8~eV, which reproduce well the
corresponding experimental values, i.e., $92.04 \pm 0.05$~keV and
$5.57 \pm 0.25$ eV~\cite{8Be}.
It is known \cite{Horiuchi} that this type of
$\alpha$-$\alpha$ potential with a repulsive core
is not suitable for describing the $0^+_1$ or $2^+_1$ state of
$^{12}$C, in which the three-$\alpha$ particles are closely
bound. However, it can successfully be applied to the $0^+_2$ state,
in which the three-$\alpha$'s are loosely coupled.
Therefore, we consider that the simulation of the
Pauli principle by introducing the repulsive part is
justified in describing the three-$\alpha$ scattering states.
In fact, it is numerically confirmed that even if we put
$v^{\rm nucl}(x)=0$ in the calculation of $V_{ii'}\left(  R\right)$
given by  Eq.~(\ref{FF}), the resulting reaction rate
$\langle \alpha\alpha\alpha\rangle (T)$ for $T\le 10^8$~K
changes by only about 2\% at most.

Equations (\ref{CDCCeq}) are numerically integrated up to
$R_{\rm max}=2,500$ fm with $\Delta R= 0.25$ fm, and
$\hat{\chi}_{i}^{(i_0)}(R)$ ($i=1$--$i_{\rm max}$)
are connected to usual asymptotic form.
The total energy $E$ is varied from 1~keV to 500~keV with $\Delta E=1$~keV;
around the Hoyle resonance at
$E=379.5$~keV, we put $\Delta E = 0.1$~keV.
In the evaluation of the coupling potentials
$V_{ii'}\left(  R\right)$
given by  Eq.~(\ref{FF}), we reduce
$v^{\rm nucl}$ by 1.5\% so that the
($\alpha_{1}$-$\alpha_{2}$)-$\alpha_3$ system in the $i=86$ channel,
with $i_0=86$, forms a resonance at $\epsilon_3=287.5$~keV.

In the calculation of $\Psi_{M}^{2^+}$, use of the
potential of Eq.~(\ref{Vaa}) is not appropriate
as mentioned above.
Instead, we adopt a sophisticated three-$\alpha$
wave function~\cite{Hiyama,GEM}
that was obtained on the basis of
the orthogonality condition model
\cite{Saito} for describing the Pauli principle
with the three-$\alpha$ particles symmetrized.
In this semi-microscopic calculation,
the $\alpha$-$\alpha$ potential was derived
by folding an effective nucleon-nucleon force
\cite{Hasegawa} into the $\alpha$-particle density.

%
\begin{figure}[b]
\begin{center}
 \includegraphics[width=0.65\textwidth,clip]{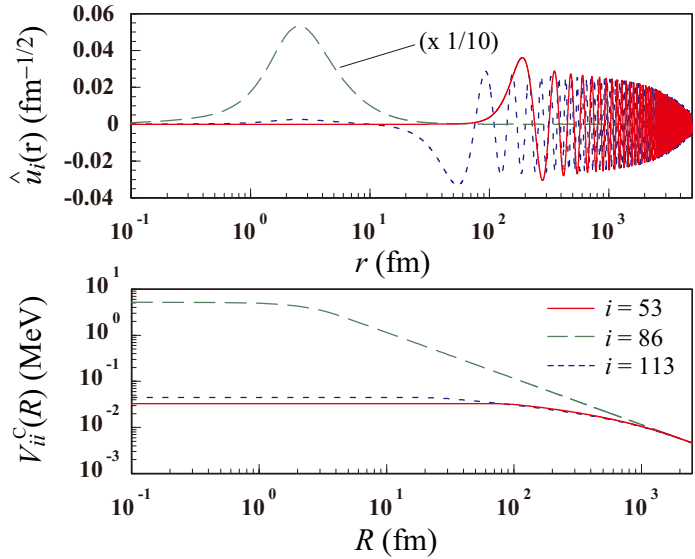}
 \caption{
 (Color online)
 Discretized  $\alpha_1$-$\alpha_2$
continuum wave functions $\hat{u}_{i}(r)$
(upper panel) and
 the diagonal Coulomb potentials $V_{ii}^{\rm C}(R)$ (lower panel).
 The solid, dashed, and dotted lines
 correspond to $i=53$, 86 (resonance), and 113, respectively.}
\label{fig3}
\end{center}
\end{figure}
We show in Fig.~\ref{fig3} the discretized continuum wave functions
$\hat{u}_{i}(r)$ of the $\alpha_1$-$\alpha_2$ system (upper panel)
and the Coulomb parts of the diagonal coupling potentials
$V_{ii}^{\rm C}(R)$ (lower panel).
We use logarithmic scale for the horizontal axis in each panel.
The solid, dashed,
and dotted lines correspond to the $\alpha_1$-$\alpha_2$
continuum states with
the average energies $\hat{\epsilon}_{12,i}$ of
38.2~keV ($i=53$), 92.0~keV ($i=86$), and 152~keV ($i=113$),
respectively.
One sees that the resonant wave function
has a dominant amplitude in the interaction region ($r \la 10$ fm),
while the nonresonant wave functions have appreciable amplitudes only at
larger $r$.
This clear difference in $\hat{u}_{i}(r)$
drastically affects $V_{ii}^{\rm C}(R)$
as shown in the lower panel.
One sees that the Coulomb barrier height $V_{ii}^{\rm C}(R)$
for the $\alpha_1$-$\alpha_2$
nonresonant bins is very much lower than for the resonant bin.
Thus, $\alpha_3$ can easily approach to the
$\alpha_1$-$\alpha_2$ system when
$\alpha_1$-$\alpha_2$ is in nonresonant states.
This important feature has not been considered in preceding studies
with Nomoto's method.

%
\begin{figure}[t]
\begin{center}
 \includegraphics[width=0.65\textwidth,clip]{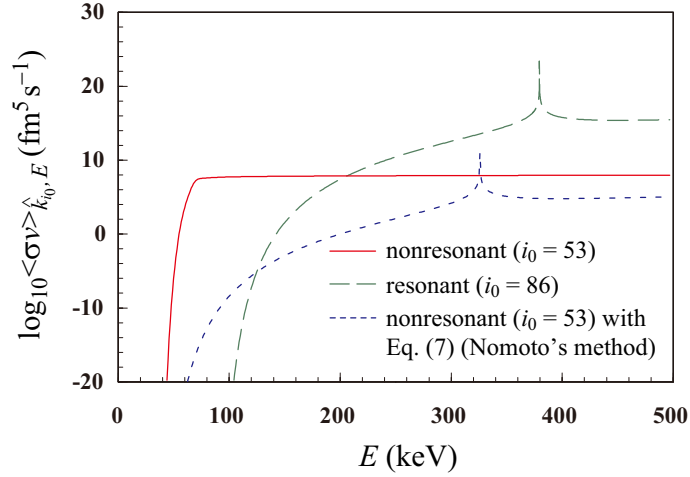}
 \caption{
 (Color online)
 $\left\langle \sigma v \right\rangle_{\hat{k}_{i_0},E}$
 for $i_0=53$ (solid line) and $i_0=86$ (dashed line). The dotted
 line shows the result for $i_0=53$ with Eq.~(\ref{rep}),
 which simulates Nomoto's method.
 }
\label{fig4}
\end{center}
\end{figure}
Figure \ref{fig4} shows the calculated results of
$\left\langle \sigma v \right\rangle_{\hat{k}_{i_0},E}$;
to show the difference between the
resonant and nonresonant results clearly,
CC effects are not included here.
The solid (dashed) line corresponds to $i_0=53$ (86).
One sees that the solid line has a completely different energy
dependence from that of the dashed line.
The nonresonant reaction probability
$\left\langle \sigma v \right\rangle_{\hat{k}_{i_0},E}$
is almost constant (in the scale of the vertical axis in Fig.~\ref{fig4})
above $E\sim 70$ keV, and dominates the resonant one for $E \la 200$~keV.
Note that for $i_0=53$, $E=70$~keV corresponds to $\epsilon_3 = 31.8$~keV
that agrees well with the {\it quenched} Coulomb barrier height shown
by the solid line in Fig.~\ref{fig3} (lower panel).

If we disregard unphysically the channel dependence of the
coupling potentials $V_{ii'}(R)$ and take only the diagonal
components, by using the following replacement:
\begin{equation}
V_{ii'}(R) \rightarrow V_{86,86}(R) \delta_{ii'},
\label{rep}
\end{equation}
we can simulate the evaluation of
$\left\langle \sigma v \right\rangle_{\hat{k}_{i_0},E}$
with Nomoto's method;
note that the 86th bin corresponds to the $\alpha_1$-$\alpha_2$
resonant state.
This replacement of $V_{ii'}(R)$ in Eq.~(\ref{CDCCeq})
makes the three-$\alpha$ system form a resonance
when $\epsilon_3=287.5$~keV, independently
of $\epsilon_{12}$.
In other words, when $\epsilon_{12}=92.0 - \Delta \epsilon_{12}$~keV,
the three-$\alpha$ system has a resonance at
$E = 379.5 - \Delta \epsilon_{12}$~keV.
This $\Delta \epsilon_{12}$ is nothing but the energy shift
in Nomoto's method.
The result for $i_0=53$ with Eq.~(\ref{rep})
is shown by the dotted line in Fig.~\ref{fig4}.
Although it has a peak at $E\sim 326$~keV as expected,
its energy dependence is similar to that of the dashed line,
which results in a much smaller value than the {\it true}
nonresonant capture probability (solid line) at low energies.
Thus, Nomoto's method is shown to be a crude approximation of the
accurate three-body model calculation.

%
\begin{figure}[t]
\begin{center}
 \includegraphics[width=0.7\textwidth,clip]{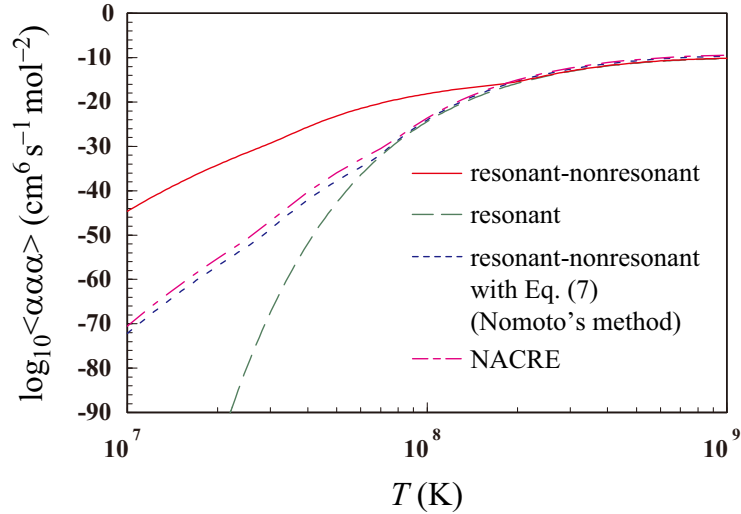}
 \caption{
 (Color online)
 Triple-$\alpha$ reaction rate as a function of
 temperature. The solid line represents the result of CDCC.
 The dashed line shows the contribution of resonant
 capture. The result of CDCC simulating Nomoto's method
 with Eq.~(\ref{rep}) is shown by the dotted line.
 The dash-dotted line shows the reaction rate of NACRE~\cite{NACRE}.
 }
\label{fig5}
\end{center}
\end{figure}
We show by the solid line in Fig.~\ref{fig5} our result of the
triple-$\alpha$ reaction rate obtained by CDCC.
The triple-$\alpha$ reaction rate of the NACRE compilation~\cite{NACRE}
is shown by the dash-dotted line for comparison. Drastic enhancement
of the reaction rate at low temperatures is found.
The dashed line shows our result including only
the resonant capture process, i.e., via the
($\alpha_{1}$-$\alpha_{2}$)-$\alpha_3$
resonance state at $E=379.5$~keV.
Difference between the solid and dashed lines
clearly shows the dominant contribution of the nonresonant triple-$\alpha$
process for $T \la 2 \times 10^8$~K.
As mentioned above, we can simulate the NACRE evaluation,
which is essentially based on Nomoto's method, by using Eq.~(\ref{rep});
the result of this calculation is shown by the dotted line.
As expected, the dotted line reproduces well the result of NACRE.
Thus, we conclude that the nonresonant capture process
has much larger contribution than
in previous evaluations~\cite{Nomoto1,Nomoto3,NACRE,Langanke},
as a result of the significant reduction of the Coulomb barrier height
between $\alpha_3$ and the nonresonant $\alpha_1$-$\alpha_2$.
This barrier reduction cannot be taken into account if one uses
Eq.~(\ref{rep}), or, equivalently, adopts Nomoto's method
as mentioned above.
Another remark is on the importance of the low-energy $\alpha_1$-$\alpha_2$
continuum states below the resonance at 92.04~keV, which were completely
missed in the preceding three-$\alpha$ model studies~\cite{Kamimura1,DB}. \
It is found that
these low-energy states have almost all contributions to the total
reaction rate (solid line) for $T\la 4.0\times 10^7$~K. Because of the lack
of these states, the reaction rate given in  Ref.~\citen{DB},
e.g., $5.34\times 10^{-63}$ cm$^6$\,s$^{-1}$\,mol$^{-2}$ at $10^7$~K,
is markedly smaller than the present result at low temperatures.

\begin{table}[htbp]
\caption{
The triple-$\alpha$ reaction rate (in cm$^6$\,s$^{-1}$\,mol$^{-2}$)
obtained by CDCC, together
with its ratio to the rate of NACRE. The number $a[n]$ means
$a \times 10^{n}$.
Rates at other temperatures are available.
}
\begin{center}
\begin{tabular}{ccccccccc}
\hline
\hline
$T$ &
$\left\langle \alpha\alpha\alpha\right\rangle$ & &
ratio &
&
$T$ &
$\left\langle \alpha\alpha\alpha\right\rangle$ & &
ratio \\
(10$^7$ K) & &  & & & (10$^7$ K) & & & \\
\hline
1   & 1.08[$-$44] & & 3.7[+26]  & &  15  & 1.52[$-$16] & & 9.5[+01]  \\
1.5 & 3.42[$-$38] & & 5.4[+23]  & &  20  & 1.92[$-$15] & & 1.9[+0]   \\
2   & 3.12[$-$34] & & 5.7[+21]  & &  25  & 4.37[$-$14] & & 1.0[+0]   \\
2.5 & 1.73[$-$31] & & 1.6[+20]  & &  30  & 4.51[$-$13] & & 9.9[$-$1] \\
3   & 2.44[$-$29] & & 1.7[+18]  & &  35  & 2.29[$-$12] & & 9.8[$-$1] \\
4   & 1.10[$-$25] & & 2.1[+15]  & &  40  & 7.37[$-$12] & & 9.8[$-$1] \\
5   & 3.41[$-$23] & & 3.3[+13]  & &  50  & 3.41[$-$11] & & 9.9[$-$1] \\
6   & 1.63[$-$21] & & 1.4[+12]  & &  60  & 8.56[$-$11] & & 9.9[$-$1] \\
7   & 2.56[$-$20] & & 8.5[+10]  & &  70  & 1.54[$-$10] & & 9.9[$-$1] \\
8   & 2.01[$-$19] & & 2.1[+09]  & &  80  & 2.26[$-$10] & & 1.0[+0]   \\
9   & 9.89[$-$19] & & 3.9[+07]  & &  90  & 2.93[$-$10] & & 1.0[+0]   \\
10  & 3.52[$-$18] & & 1.5[+06]  & &  100 & 3.48[$-$10] & & 1.0[+0]   \\
\hline
\hline
\end{tabular}
\label{tab1}
\end{center}
\end{table}
For more detailed comparison, we need renormalization
for the CDCC result.
We introduce an effective charge $\delta e=0.77 e$ to the E2
transition operator, so that our result of the B(E2) value,
evaluated at $E=379.5$~keV with the $0^+$
wave function normalized to unity,
reproduces the experimental value of 13.4 $e^2$\,fm$^4$~\cite{12C},
just in the same way as in Ref.~\citen{DB}. \
Then, we renormalize our result (with $\delta e$) to the reaction rate
of NACRE at $T=10^9$~K, where the resonant capture process through
the Hoyle resonance is dominant~\cite{NACRE};
the renormalization factor obtained is 1.54.
From this value, one may estimate the uncertainty of the present
calculation to be more or less 50\%.

In Table \ref{tab1} we show the renormalized triple-$\alpha$ reaction rate
obtained by CDCC, together with its ratio to the rate of NACRE,
at some typical temperatures. As expected, for $T \ge 2.5 \times 10^8$~K
the ratio is almost unity that shows our calculation, with the
renormalization at $10^9$~K, reproduces very well the
temperature dependence of the resonant triple-$\alpha$ process
through the Hoyle resonance.
At low temperatures, on the other hand,
the ratio exceeds $10^{20}$, which will
affect the helium burning in accreting white
dwarfs and neutron stars. Furthermore,
even at rather high temperature of $1.5 \times 10^8$~K, we obtain
a larger reaction rate by almost two orders-of-magnitude
than that of NACRE.
It should be noted that a broad $2^+_2$ resonance state at
$E=1.75$~MeV with the $\alpha$-decay width of 0.56~MeV~\cite{DB}
is included in the NACRE evaluation. In fact, it is shown
that this $2^+_2$ state has significant contribution
to the triple-$\alpha$ reaction rate for
$T \ga 2 \times 10^9$~K~\cite{NACRE}. \
At these high temperatures, the nonresonant triple-$\alpha$ process
that we focus on in the present study will obviously be
negligible compared to resonant processes via the $0_2^+$, $2_2^+$, and
other possible resonance states~\cite{Nature} of $^{12}$C.
Very recently, it was reported \cite{DP} that a stellar evolution
model computed with our new reaction rate of the triple $\alpha$
reaction caused inconsistency with observations of red giant
branches. Further investigation on this implication will be
very interesting and important.

In summary, we have evaluated the triple-$\alpha$ reaction rate
by directly solving the three-body Schr\"{o}dinger equation
with CDCC.
We treat the resonant and nonresonant processes on the same footing.
The $\alpha$-$\alpha$ continuum states below the
resonance at 92.04~keV are shown to play essential roles in the
nonresonant triple-$\alpha$ process for $T \la 4.0 \times 10^7$~K.
The key of the nonresonant capture process
is that the Coulomb barrier between the two-$\alpha$'s and the
third $\alpha$ is much quenched compared to that in the resonant
capture.
This property extremely enhances the nonresonant
triple-$\alpha$ process at low temperatures, i.e., $T \la 10^8$~K.
The ratio of our triple-$\alpha$ reaction rate to that of the NACRE
compilation is more than $10^{26}$ at $10^7$~K and about 100
at $1.5 \times 10^8$~K.
It is found that Nomoto's method for three-body
nonresonant capture processes,
which is used in the NACRE compilation and many other studies,
is a crude approximation, with Eq.~(\ref{rep}),
of the accurate quantum three-body model calculation.
The newly evaluated triple-$\alpha$ reaction rate will affect
many studies on
nuclear astrophysics, those on helium burning in accreting white
dwarfs and neutron stars in particular.
Detailed description of the theoretical framework together
with further discussion on the comparison with other existing
methods will be presented in a forthcoming paper.

\vspace{3mm}

The authors wish to thank M.~Kawai, M.~Hashimoto, and E.~Hiyama
for helpful discussions.


\end{document}